# OPTIMIZATION OF PATTERNED SURFACES FOR IMPROVED SUPERHYDROPHOBICITY THROUGH COST-EFFECTIVE LARGE-SCALE COMPUTATIONS

**V. Krokos[1], G. Pashos[1,*], A. N. Spyropoulos[1], G. Kokkoris[1,2], A. G. Papathanasiou[1], A. G. Boudouvis[1]**

[1]School of Chemical Engineering, National Technical University of Athens, Zografou Campus, Athens 15780, Greece

[2]Institute of Nanoscience and Nanotechnology, NCSR «Demokritos», Athens 15310, Greece

**ABSTRACT**

The growing need for creating surfaces with specific wetting properties, such as superhyrdophobic behavior, asks for novel methods for their efficient design. In this work, a fast computational method for the evaluation of patterned superhyrdophobic surfaces is introduced. The hydrophobicity of a surface is quantified in energy terms through an objective function. The increased computational cost led to the parallelization of the method with the Message Passing Interface (MPI) communication protocol that enables calculations on distributed memory systems allowing for parametric investigations at acceptable time frames. The method is demonstrated for a surface consisting of an array of pillars with inverted conical (frustum) geometry. The parallel speedup achieved allows for low cost parametric investigations on the effect of the fine features (curvature and slopes) of the pillars on the superhydophobicity of the surface and consequently for the optimization of superhyrdophobic surfaces.

## 1. Introduction

The wetting properties of a surface crucially affect its suitability for a variety of applications including self-cleaning[1], resistance to corrosion[2] and microfluidics[3]. The wetting properties of surfaces can be modified through thermal[4], chemical[5], or electric[6] means. However, one of the most common ways to alter the wetting properties of a flat surface is to pattern it[7]. The most common patterns are pillar arrays. The geometric characteristics of the pillars significantly affect the wetting properties of the surfaces.

Droplets on top of patterned surfaces will be in either the Cassie-Baxter (CB) state or the Wenzel (W) state. The CB state is characterized by reduced contact of the droplet with the solid and free movement with low friction between the droplet and the surface[8,9]. In contrast to that, the W state is characterized by pinning of the liquid on the geometric features of the pillars[8,9]. It is obvious that the desired state for the liquid is the CB. However, this state is metastable and can easily transition to W, which is undesirable. The resistance to the transition is indicative of the hydrophobicity of a patterned surface and thus it is used as an objective function.

The transition from the CB to the W state requires a minimum amount of energy (energy barrier), the same way that a chemical reaction has an activation energy. The energy barrier is calculated through the Minimum Energy Paths (MEPs) that reveal the course of the CB-W transition. The energy barrier is the objective function for the evaluation of the superhydrophobicity of the patterned surface.

* Corresponding author gpashos@chemeng.ntua.gr

The energy barriers are sensitive to small changes in the geometry. This feature allows to investigate the effects of not only the main shape (e.g. rectangular, cylinder, T-shaped pillars) and size of the pillars but also of the finer geometric characteristics of the pillars (e.g. multi-scale patterns on the main pillars, curvature radius). The focus of this work is on an array of pillars with an inverted conical (frustum) geometry[10] carrying several finer geometric characteristics.

The MEPs can reveal, apart from the energy barriers, additional information about the CB-W transitions. The transitions may occur in many different ways, however they may be categorized in a few distinct transitions, commonly referred to as failure modes. The study of the failure modes can lead to improved design of patterns for improved superhydrophobic behavior. Two such cases are demonstrated in this paper, T* failure and H* failure.

The MEPs are calculated through an iterative method namely the String Method. Despite the many advantages that MEPs offer, their calculation comes with a significant computational cost. This problem is tackled with the efficient parallelization of the code using the Message Passing Interface (MPI) communication protocol allowing for High Performance Computing (HPC) on distributed memory systems. Moreover, the mathematical formulation of the problem allows for investigation of a 3D physics problem with a 2D mathematical/numerical grid, massively reducing the computational cost. Two different computational domains are proposed, both taking advantage of the symmetry of the problem at different levels. The small computational domain is suitable for an initial assessment of pillars and later on a full investigation can be performed on the big computational domain for more detailed examination.

There are several methods for the prediction of the shape of the air-liquid interface such as lattice Boltzmann[11,12], molecular dynamics[13,14] and phase-field[15-17] but in this work we chose a more classical approach, the Young-Laplace equation. Even though Young-Laplace equation has a reduced computational cost compared to the above mentioned methods, when implemented for complex geometries it can be challenging and impractical. In order to tackle this problem a new equation is formed, the modified Young-Laplace Equation, by adding an extra term which allows for easy manipulation of all types of surface patterns[18]

## 2. Mathematical Formulation

### 2.1. Modified Young-Laplace Equation

The Young-Laplace equation (1) correlates the pressure difference across a liquid-liquid interface with the shape of that interface. The modified Young-Laplace Equation (2) is formed by adding the term, $P_{LS}$, to (1).

$$P_L = 2H \tag{1}$$

$$2H + P_{LS} = P_L \tag{2}$$

Where $P_L$ is the dimensionless pressure difference across the fluid interface (Laplace Pressure), $H$ is the local mean curvature of the interface, $P_{LS}$ is the dimensionless pressure that encapsulates the liquid-solid interactions.

This expression for the $P_{LS}$ (3.1) is based on Smoothed Particle Hydrodynamics (SPH)[19] models. It is a simple polynomial expression with a single parameter $\delta$, the dimensionless distance from the solid boundary. $P_{LS}$ is positive when $\delta<0$ creating repulsive forces between liquid and solid when the first enters the second and also $P_{LS}$ is positive when $\delta>0$ creating attractive forces between the liquid and the solid when the first is inside the range of liquid-solid interactions. $w_{LS}$ is the wettability of the solid (dimensionless magnitude of the liquid solid interactions), $h$ is the dimensionless maximum distance where liquid-solid interactions take place. When $\delta>h$ then $P_{LS}$ is considered zero because no interactions take place between the liquid and the solid. Finally $U$ is the dimensionless potential of the interface.

$$P_{LS} = w_{LS} (\delta - h)^3 \delta^3, \delta<h \tag{3.1}$$

$$P_{LS} = -\frac{dU}{d\delta} \tag{3.2}$$

### 2.2. Dynamic Equations

Even though Modified Young-Laplace Equation (2) is the governing equation, implementing the String Method requires a dynamic equation (4) for the prediction of equilibrium states. This equation doesn't have to be necessarily exact. Based on that, we created a pseudo-transient scheme[18] (Figure 1). For simplicity the interface is moving at the normal direction and the velocity is proportional to the mean curvature plus the $P_{LS}$ minus the $P_L$.

$$\frac{d\mathbf{r}}{dt} = \mu(2H\mathbf{n} + P_{LS}\mathbf{n} - P_L\mathbf{n}) \tag{4}$$

Where $\mathbf{r}$ is the position vector of the interface, $\mu$ is the mobility of the interface that is considered 1 for this study and $\mathbf{n}$ is the unit normal vector to the interface. The steady-state solution of (4) regardless of the initial conditions and the employed dynamics satisfies (2).

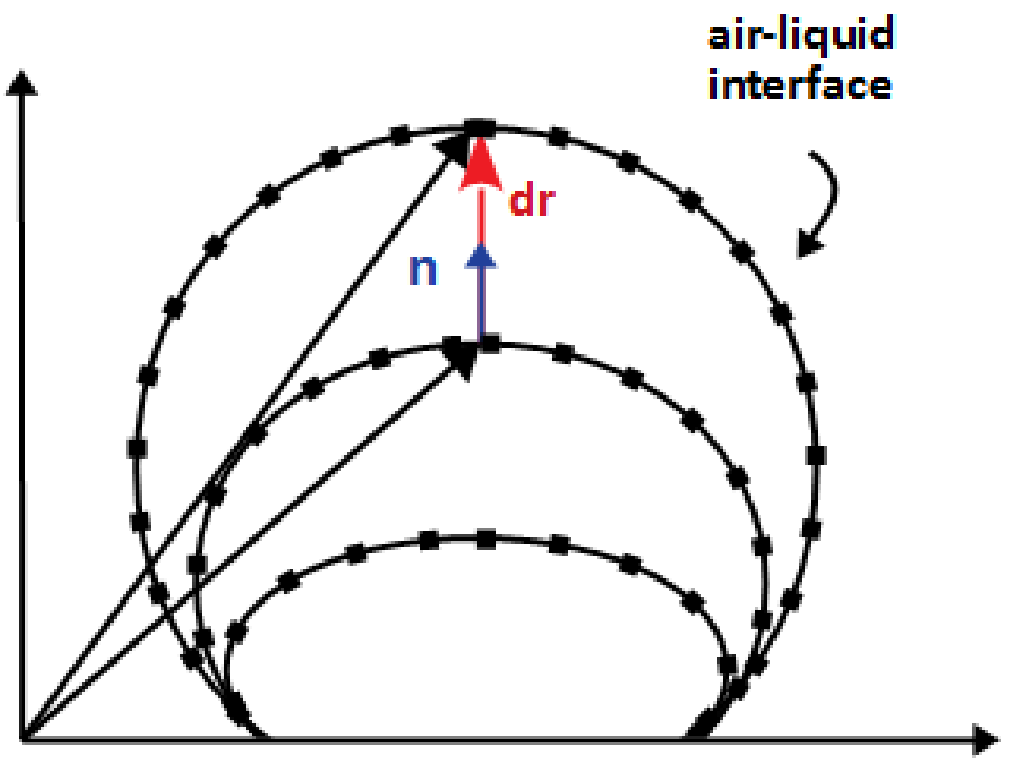

*Figure 1: Pseudo-transient scheme of the Modified Young-Laplace equation*

### 2.3. Boundary Value Problem (BVP)

The governing equation predicts the shape of the 3D air-liquid interface. The shape and the dimensionality of the interface are determined by the 3 dependent spatial variables x = x(u,v), y = y(u,v), z = z(u,v), where x, y, z is the Cartesian coordinates of the interface and u, v are independent variables $u,v \in [0,1]$ . Based on the above formulation the 3D interface is parameterized as follows: $\mathbf{r}=\mathbf{r}(x(u,v),y(u,v),z(u,v))$ . In order to incorporate the dynamic equation (4) into the problem, the mean curvature has to be expressed in terms of *u*, *v*. Based on the first and second fundamental forms of differential geometry the mean curvature can be described with equation[20] (5)

$$2H=\frac{eG-2fF+\mathrm{gE}}{EG-F^2} \tag{5}$$

Where

$$E=\mathbf{r}_u\cdot\mathbf{r}_u \quad , \quad G=\mathbf{r}_v\cdot\mathbf{r}_v \quad , \quad F=\mathbf{r}_u\cdot\mathbf{r}_v \quad , \quad e=\mathbf{r}_{uu}\cdot\mathbf{n} \quad , \quad g=\mathbf{r}_{vv}\cdot\mathbf{n} \quad , \quad f=\mathbf{r}_{uv}\cdot\mathbf{n}$$

$$\mathbf{r}_u=\frac{\partial\mathbf{r}}{\partial u} \quad , \quad \mathbf{r}_v=\frac{\partial\mathbf{r}}{\partial v} \quad , \quad \mathbf{r}_{uu}=\frac{\partial^2\mathbf{r}}{\partial u^2} \quad , \quad \mathbf{r}_{vv}=\frac{\partial^2\mathbf{r}}{\partial v^2} \quad , \quad \mathbf{r}_{uv}=\frac{\partial^2\mathbf{r}}{\partial u\partial v}$$

Where the components of the unit normal vector $\mathbf{n}$ are:

$$n_x=\frac{y_u z_v-y_v z_u}{\sqrt{EG-F^2}} \quad , \quad n_y=\frac{x_v z_u-x_u z_v}{\sqrt{EG-F^2}} \quad , \quad n_z=\frac{x_u y_v-x_v y_u}{\sqrt{EG-F^2}}$$

Where *E*, *F*, *G* are the coefficients of the first fundamental form and *K*, *L*, *M* are the coefficients of the second fundamental form.
The combination of (4) and (5) yields the following system of PDEs.

$$\frac{\partial x}{\partial t}=\frac{Gx_{uu}+Ex_{vv}-2Fx_{uv}}{EG-F^2}+P_{LS}n_x-P_L n_x \tag{6.1}$$

$$\frac{\partial y}{\partial t}=\frac{Gy_{uu}+Ey_{vv}-2Fy_{uv}}{EG-F^2}+P_{LS}n_y-P_L n_y \tag{6.2}$$

$$\frac{\partial z}{\partial t}=\frac{Gz_{uu}+Ez_{vv}-2Fz_{uv}}{EG-F^2}+P_{LS}n_z-P_L n_z \tag{6.3}$$

The equations are discretized using the backward Euler method. The forward Euler method is simpler but very unstable for this case. It requires a very small time step, rendering its implementation much slower than the backward Euler. For the spatial discretization a central difference scheme is implemented, for improved numerical stability.

The shape of the air-liquid interface is investigated in a cube with a patterned surface. Those problems usually involve a droplet on top of thousands pillars but assuming symmetric wetting, we consider only a part of the droplet on top of a pattern of 6×6 pillars (Figure 2) in order to reduce the computational cost. For an investigation of the finer geometric features of the pillars, symmetry can be additionally exploited and calculations can be performed only at one quarter of the pillar rather than the whole domain (Figure 3). Most wetting problems involve either a 2D droplet that is assumed to be infinitely stretched along one axis[17] for reduced computation cost but also limited accuracy or a 3D droplet[11] for more detailed examination but with a significantly increased computational cost. The proposed mathematical formulation allows for the parameterization of a 3D surface (3 dependent variables) using only 2 independent variables u, v thus it only requires a 2D grid for the prediction of the 3D surface. This results in great accuracy while keeping the computational load at affordable levels.

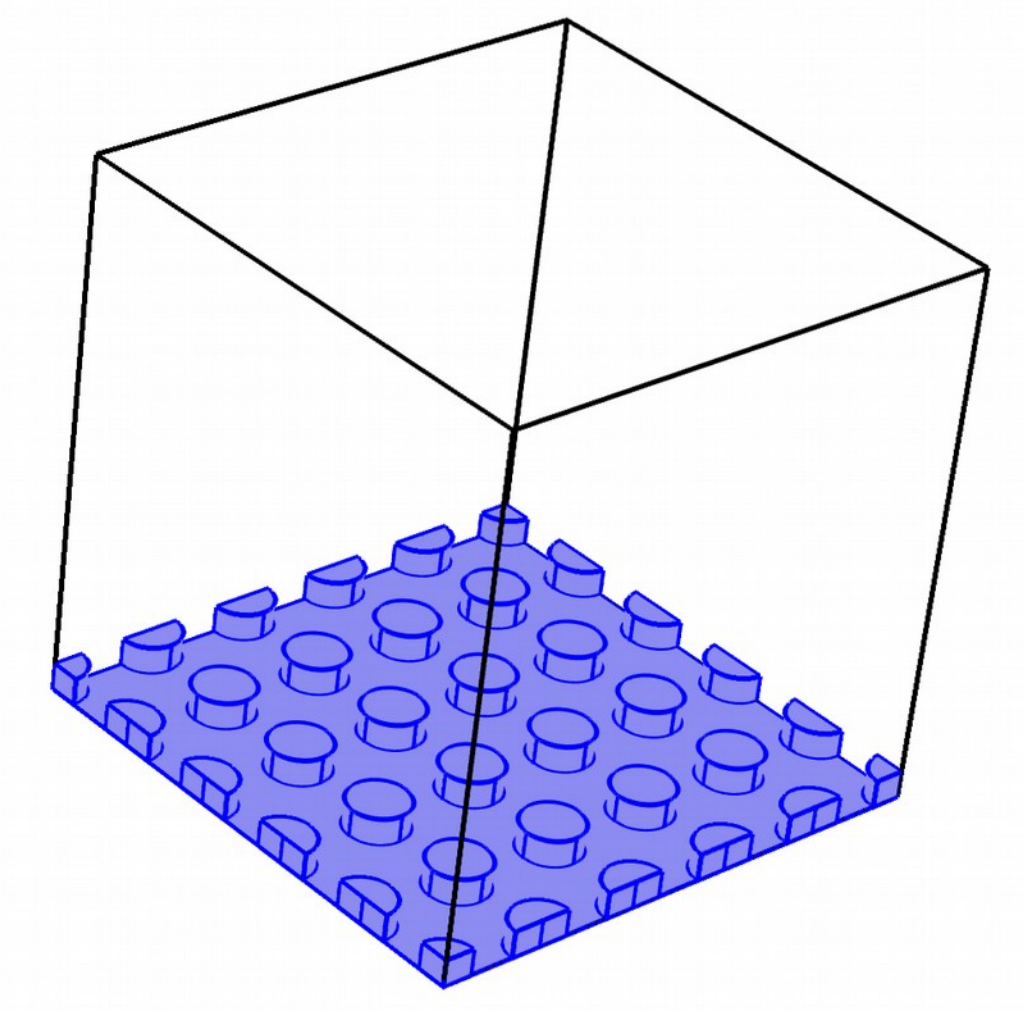

Figure 2: Unit cube, with patterned surface

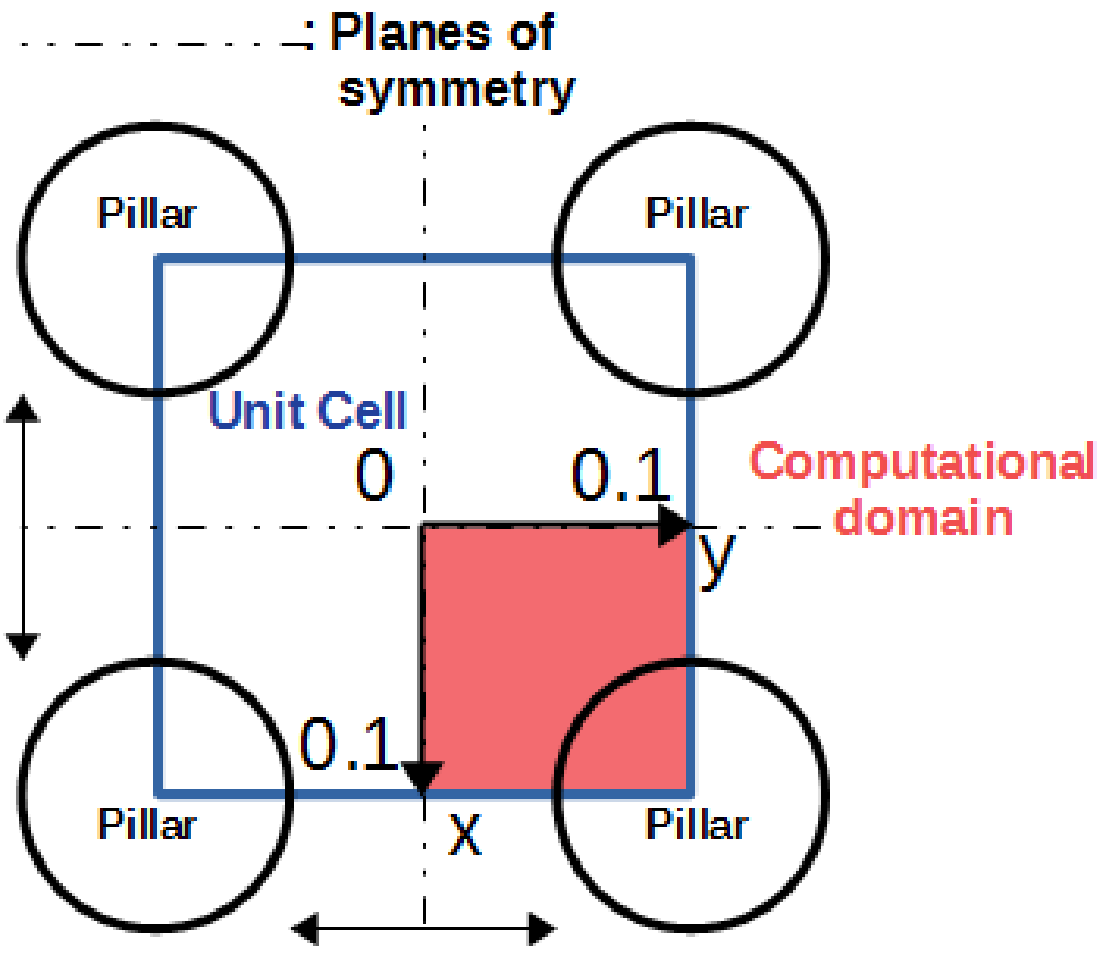

Figure 3: Reduced computational domain

The computational domain is the unit square $u, v \in [0,1]$ and there are 3 boundary conditions for every side of the square (Figure 4).

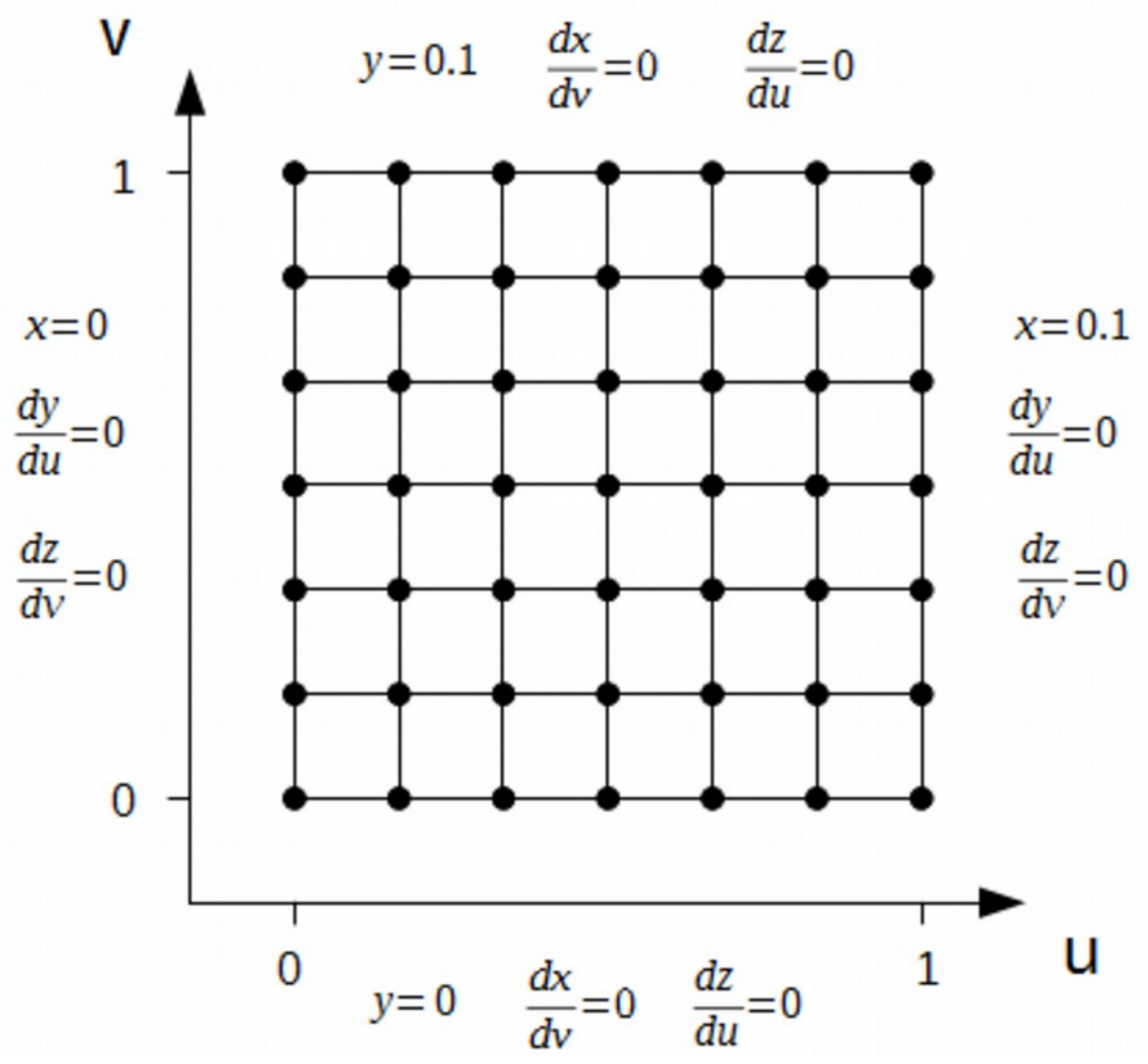

Figure 4: Computational grid with boundary conditions

## 2.4. Linear Solver

The nonlinear system that is created on every time step is solved using the Newton method. Newton actually creates a linear system of equations that is solved by implementing the Biconjugate Gradient Stabilized Method (BICGSTAB)[21]. BICGSTAB is the evolution of the Biconjugate Gradient Method (BiCG)[22] that presents smoother and faster convergence behavior. BICGSTAB belongs to the Krylov subspace methods[23] along with IDR(s) (Induced Dimension Reduction), GMRES (Generalized Minimum Residual), QMR (Quasi Minimal Residual), TFQMR (transpose-free QMR), and MINRES (minimal residual) methods. BICGSTAB in contrast with BiCG is used for solving nonsymmetric linear systems and is actually a combination of BiCG and GMRES[24,25], where every step of the BiCG is followed by one GMRES step in order to correct the irregular convergence of the BiCG. Due to the large size of the system, matrix sparsity is exploited to avoid unnecessary computations.

## 2.5. Eikonal Equation

The effect of pillar geometry is the main factor investigated. The geometry of the pillars is indirectly entered in the problem through the parameter $\delta$, the distance from the solid boundary. This distance is calculated by the Eikonal equation. The solution of the Eikonal equation is very fast for simple geometries (e.g. square, cylinders etc.) but can be very computationally intense for more complex geometries that involve non-monotonic curves and multi-scale patterns[26]. The solution of the Eikonal equation doesn't add a considerable computational load to the overall method because it is solved only once at the beginning of the method. The solution of the Eikonal equation in finite elements packages returns the distance of every point of an unstructured mesh from the solid boundary. The solution is extrapolated on a uniform orthogonal mesh and is provided for the implementation of the string method. During the calculations the distance of points that do not coincide with the nodes of the mesh is calculated through fast interpolation routines such as cubic spline interpolation method.

# 3. String Method

The modified Young-Laplace equation can predict the equilibrium wetting states. Nevertheless, the wetting states alone can't determine the suitability of the pillars for applications that require superhydrophobic behavior. For that purpose, the String Method is used in order to quantify the superhydrophobicity of the patterned surface.

Specifically, the resistance to the CB-W transition is going to be calculated in energy terms. In the free energy landscape the CB and W, which are stable equilibrium states, are local minima[27,28]. The course of the transition follows a path of least resistance that is called Minimum Energy Path (MEP) [29,30]. The MEP is aligned with the energy valleys of the free energy landscape (CB and W) and thus it passes through the corresponding energy saddle (unstable equilibrium state)[31]. The resistance to CB-W transition is quantified by the energy barrier; the latter is calculated as the difference of free energy between the saddle point and the CB state (energy barrier)[32,33]. Consequently, the objective function is the energy barrier of the CB-W transition (Figure 5). High energy barriers mean high resistance to the CB-W transition and thus enhanced superhydrophobic behavior. The CB-W transition can have multiple energy saddles, in that case the total energy barrier is equal to the cumulative sum of the energy barriers of every transition from a minima to a saddle.

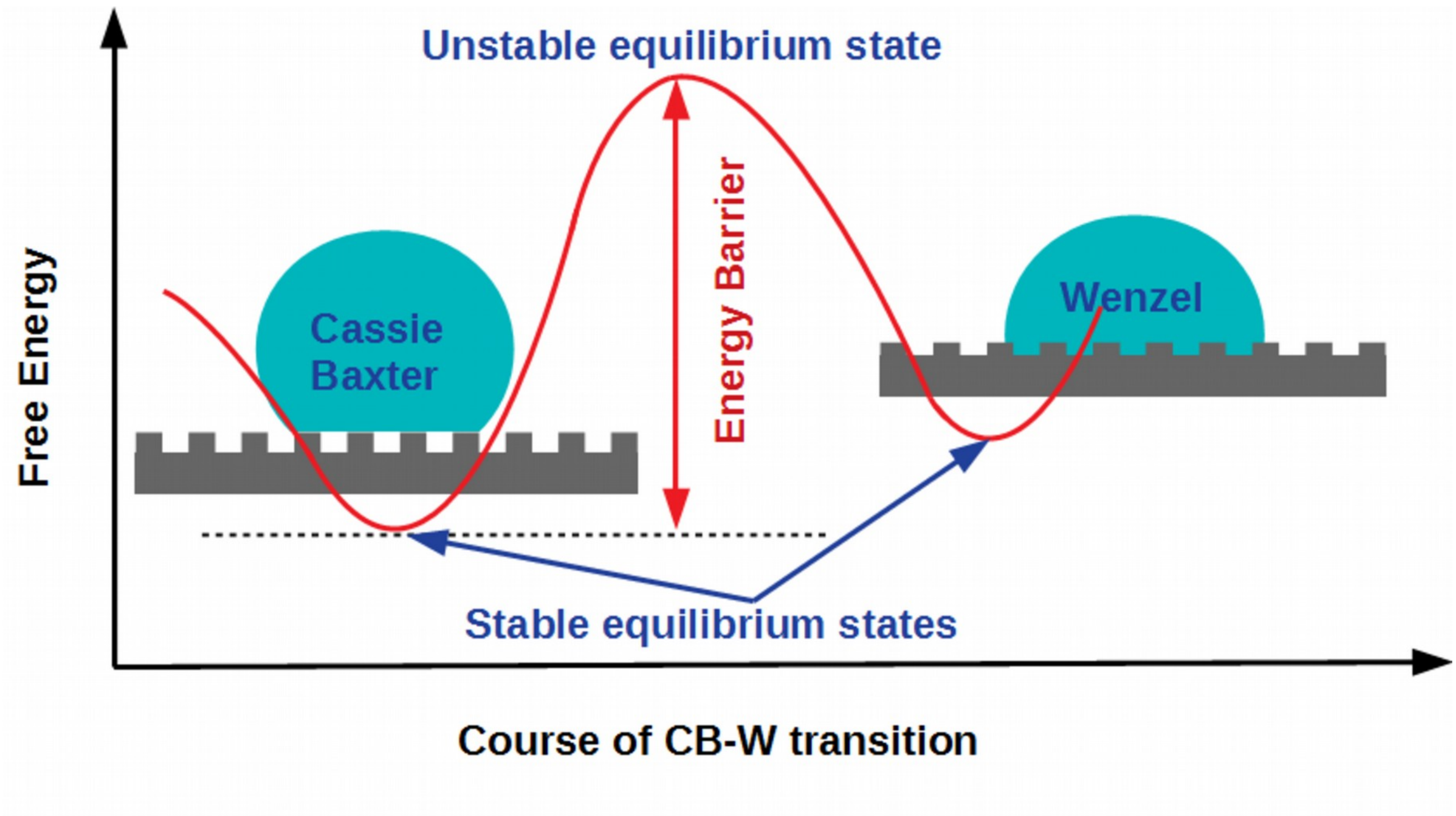

*Figure 5: Representation of a MEP connecting the stable equilibrium states (CB and W). The energy barrier of the transition is shown.*

A MEP is actually a smooth curve on the M-dimensional space, where M is equal to the degrees of freedom (DOFs) of the discretized BVP. The MEPs are approximated by a finite number of points connected by straight lines. The greater the number of these points the greater the accuracy of the method but also the computation cost. Each of these points is called Image. Every Image is the energy of a system described by the equations (6.1), (6.2), (6.3). The energy of the system is calculated by taking into account the energy of the air-liquid interface of the parameterized surface (first term of (7)) and the contribution of the liquid-solid interactions (second term of (7)). The energy is dimensionless and can only be compared with other energies calculated with the same method. The energy quantities are dimensionless as they are derived from the pressure terms, $P_L$ and $P_{LS}$, which are also dimensionless. The pressure terms are obtained in physical units by dividing with the term $l_0/\gamma_{LV}$, where $l_0$ is characteristic length equal to width of the solid surface and $\gamma_{LV}$ is the liquid-vapor interfacial tension.

$$E=\int_0^1\int_0^1\sqrt{EG-F^2}\,dudv+\int_0^1\int_0^1 U\sqrt{EG-F^2}\,dudv \tag{7}$$

The String Method can be used in order to calculate these curves[34,35]. An initial curve is used with Images starting from the CB state and reaching the W state. The string method involves one time step for the equations (6.1-3) followed by a reparametrization step. After each time step this curve will converge to the MEP. Without the reparametrization step, all the Images would unavoidably head towards an energy minimum. That would determine the CB and W states but not the MEPs that are essential for the calculation of the energy barriers. The reparametrization step ensures that all consecutive Images will have equal distance and thus they will reveal the whole MEP.

## 4. Results

### 4.1. Parallelization

Even for the small problem, where only one quarter of the pillar is investigated, the computational cost is considerable. Three equations are solved on every node of the computational mesh and assuming a 50x50 uniform grid and 40 Images for the representation of the MEP, the Degrees Of Freedom (DOFs) of the discretized BVP are DOFs = 50x50x40x3 = 300.000. This calculation needs to be repeated for at least 100.000 time steps and it takes several hours on a single core (Xeon CPUs X5660 @ 2.80GHz), rendering a full parametric investigation very demanding. That is one of the biggest obstacles when trying to conduct computational studies for investigation of interfacial phenomena. To overcome this problem the MPI protocol was implemented in order to parallelize the original code, allowing for calculations on distributed memory systems using several cores thus making our calculations extremely cost-effective.

There are many methods for the parallel solution of PDEs such as domain decomposition[36] and master-slave scheme[37]. However, in the present work instead of the PDEs the string method is parallelized by dividing the Images among the cores and so each core has all the necessary information needed to solve the PDEs (Figure 6). That results in limited communication between the cores. Communication is only performed during the reparametrization step (Figure 7).

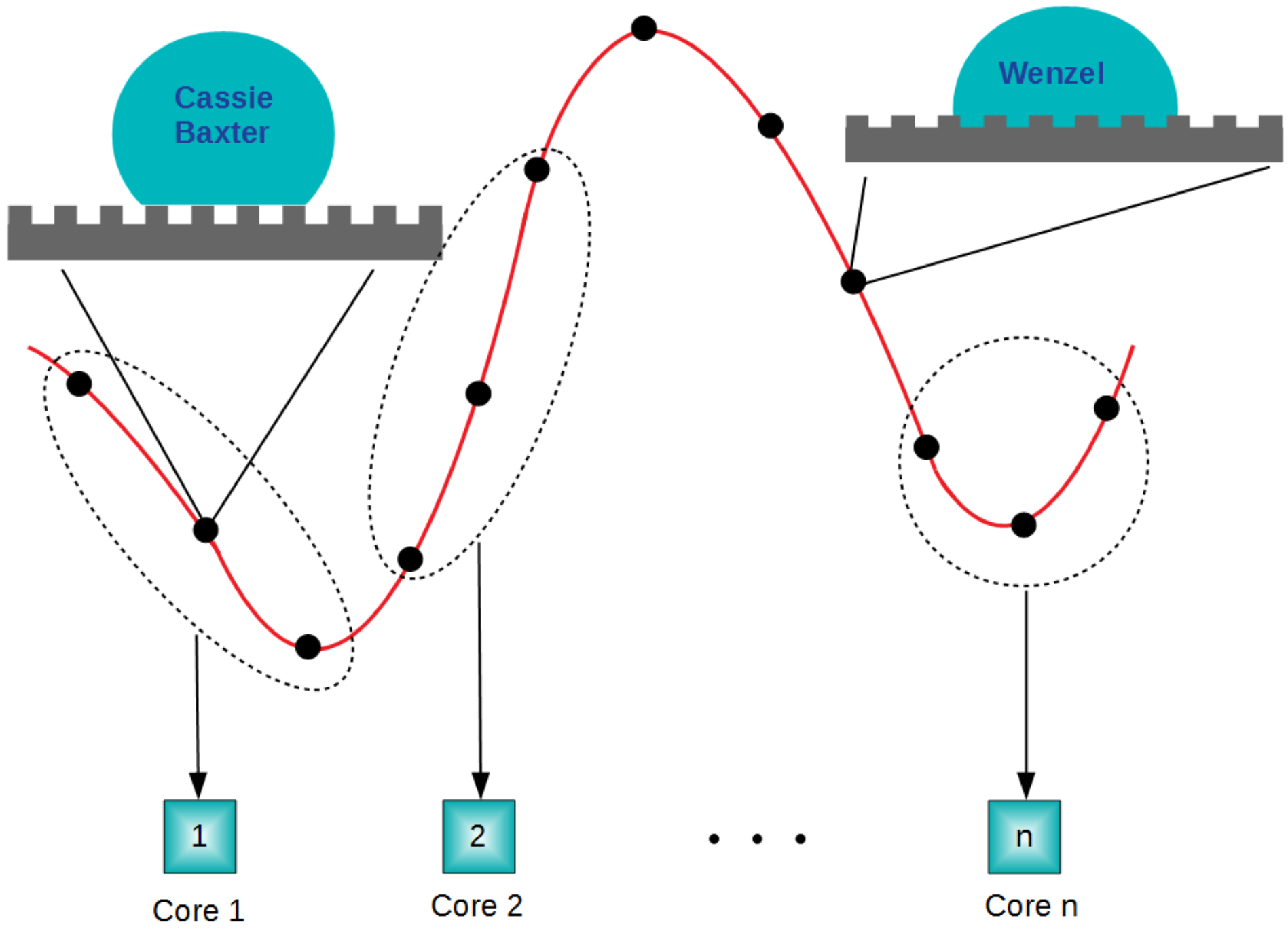

*Figure 6: Images divided among the cores*

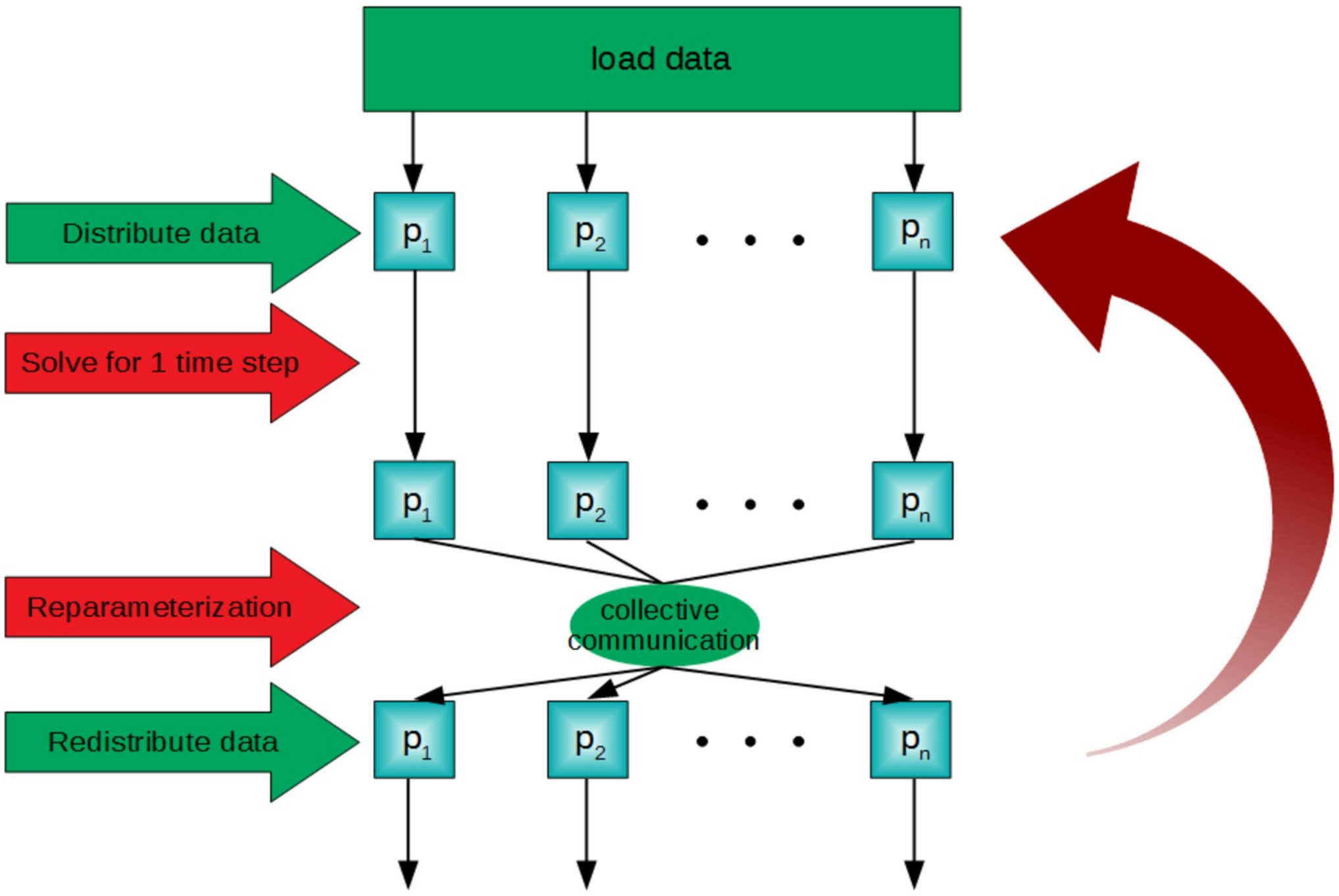

*Figure 7: Overview of the string method*

All the computations took place in the "Andromeda" computational cluster, of the School of Chemical Engineering of the National Technical University of Athens (NTUA)[38]. Andromeda is a 4 node computational cluster, with 8 Xeon CPUs X5660 @ 2.80GHz, 48 cores and 64 GB RAM. The nodes are interconnected with a Gigabit Ethernet network. Communication between cores on different nodes is much slower than communication between cores on the same node due to the relatively slow Gigabit Ethernet network. This results in increase of the communication cost with increase of the used nodes.

The efficiency of the parallel code is investigated based on the speedup. The speedup (S) is defined as the ratio of the sequential time ($t_S$) and the parallel time ($t_P$), $S = t_S/t_P$, for a given problem. The speedup is going to be investigated on problems of different mesh size and Image number. The second case seems more promising because every core will be responsible for several Images.

First we examine the speedup across different mesh sizes (Figure 8). The investigated meshes are of size 50x50 (blue line), 100x100 (green line) and 175x175 (red line) and 40 Images are used for the representation of the MEP. An almost linear speedup is observed for all mesh sizes when the number of cores is less or equal to 20. On the other hand, striking differences emerge when 40 cores are used. This is attributed to the relative increase of the communication cost in every case. Each node consists of 12 cores, so when using 20 cores only 2 nodes are required (10 cores on each node), whereas 40 cores require the use of all the 4 nodes, significantly contributing to the increase of the communication cost. The blue line, corresponding to a coarse mesh suited for calculations only on one quarter of the geometry, presents a mild slowdown. That happens because the communication cost overcomes the time spend on actual calculations. As the size of the mesh increases so does the communication cost and the time spent inside the solver. However, the time spend inside the solver increases much faster than the time spend for the communication (Figure 9). This results in improved speed up as the size of the mesh increases, green line and red line.

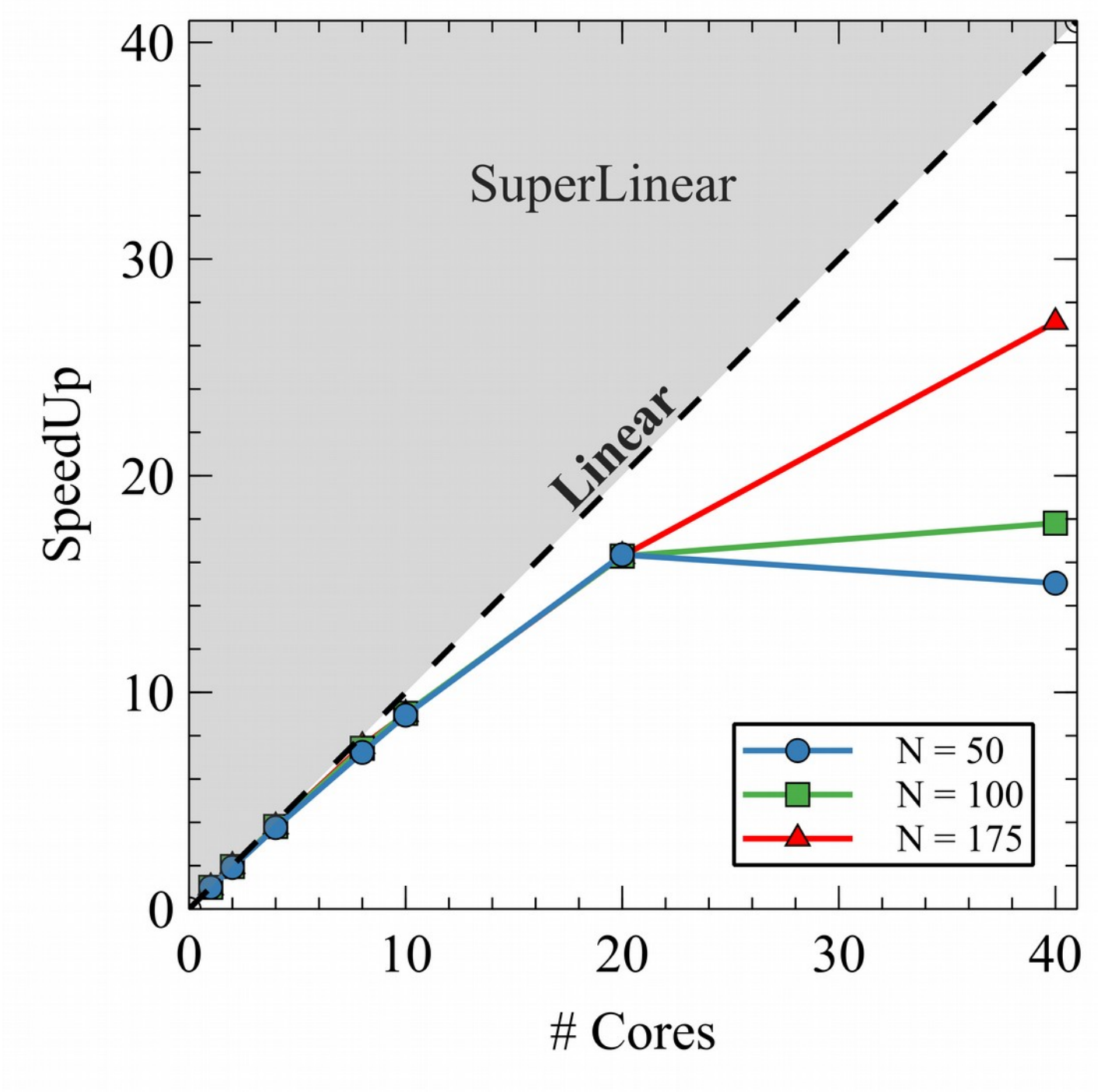

Figure 8: Speedup for different meshes of size NxN

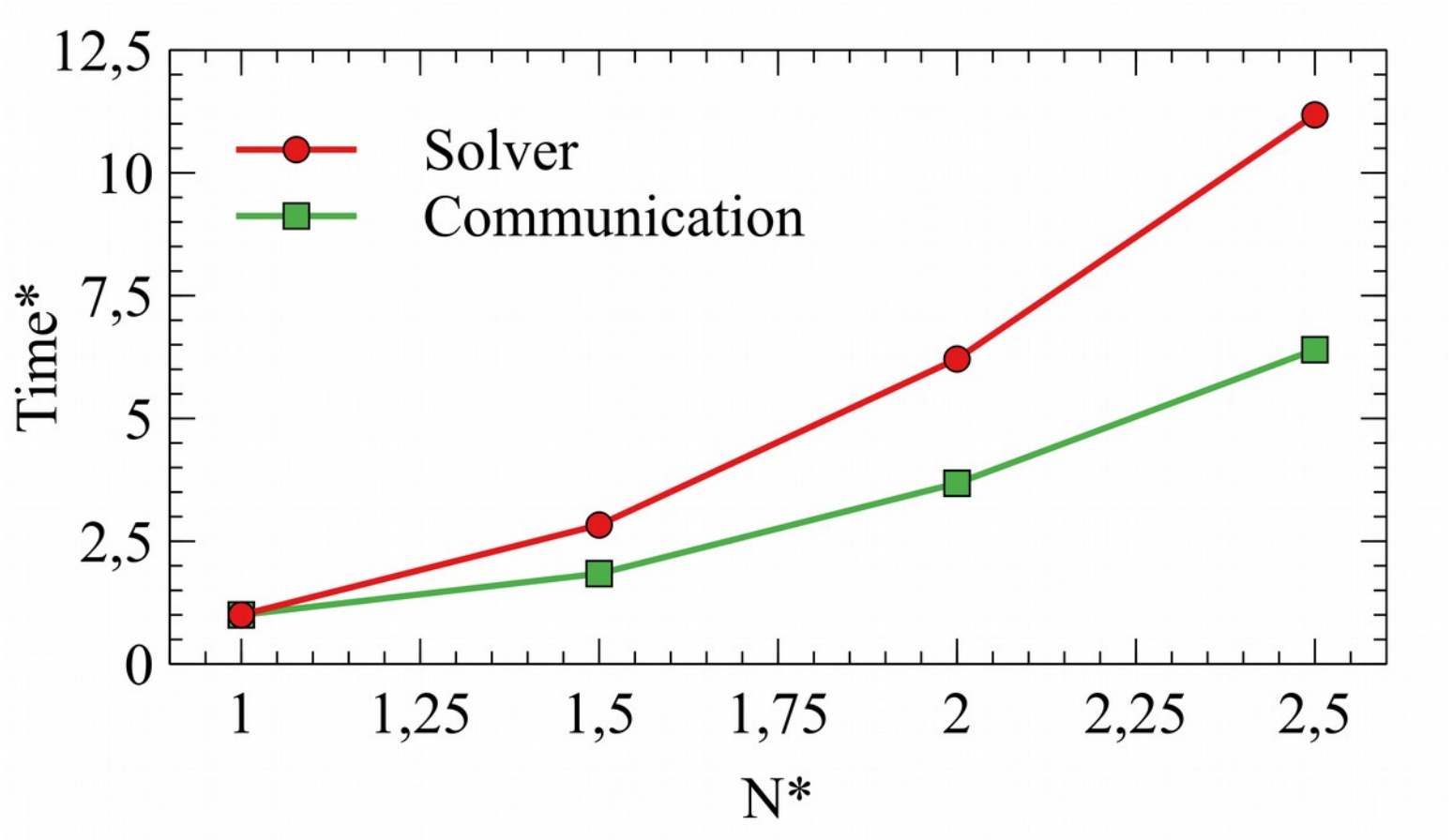

Figure 9: Diagram depicting the time needed for the solution of the system and the communication with respect to the mesh size. N* denotes the relative growth of a mesh with inital size 50 x 50 and Time* the corresponding time

Calculations on the full computational domain, 6x6 array of pillars, require both a denser mesh and an increased number of Images. Three number of Images are investigated 40, 160, 400 and a mesh of size 125x125, more suitable for the full computational domain (Figure 10). Again, when using up to 20 cores the speedup is almost linear and worth mentioning differences occur only when using 40 cores. In all the three cases the only thing that changes is the number of Images every core is responsible for. When using 40 cores this number is 1, 4 and 10 respectively for the three study cases. For the blue line, 40 Images, no further speedup is observed from 20 to 40 cores. On the contrary, when increasing the number of Images per core, green and red line, the speedup is significantly improved, for the same reasons as discussed earlier.

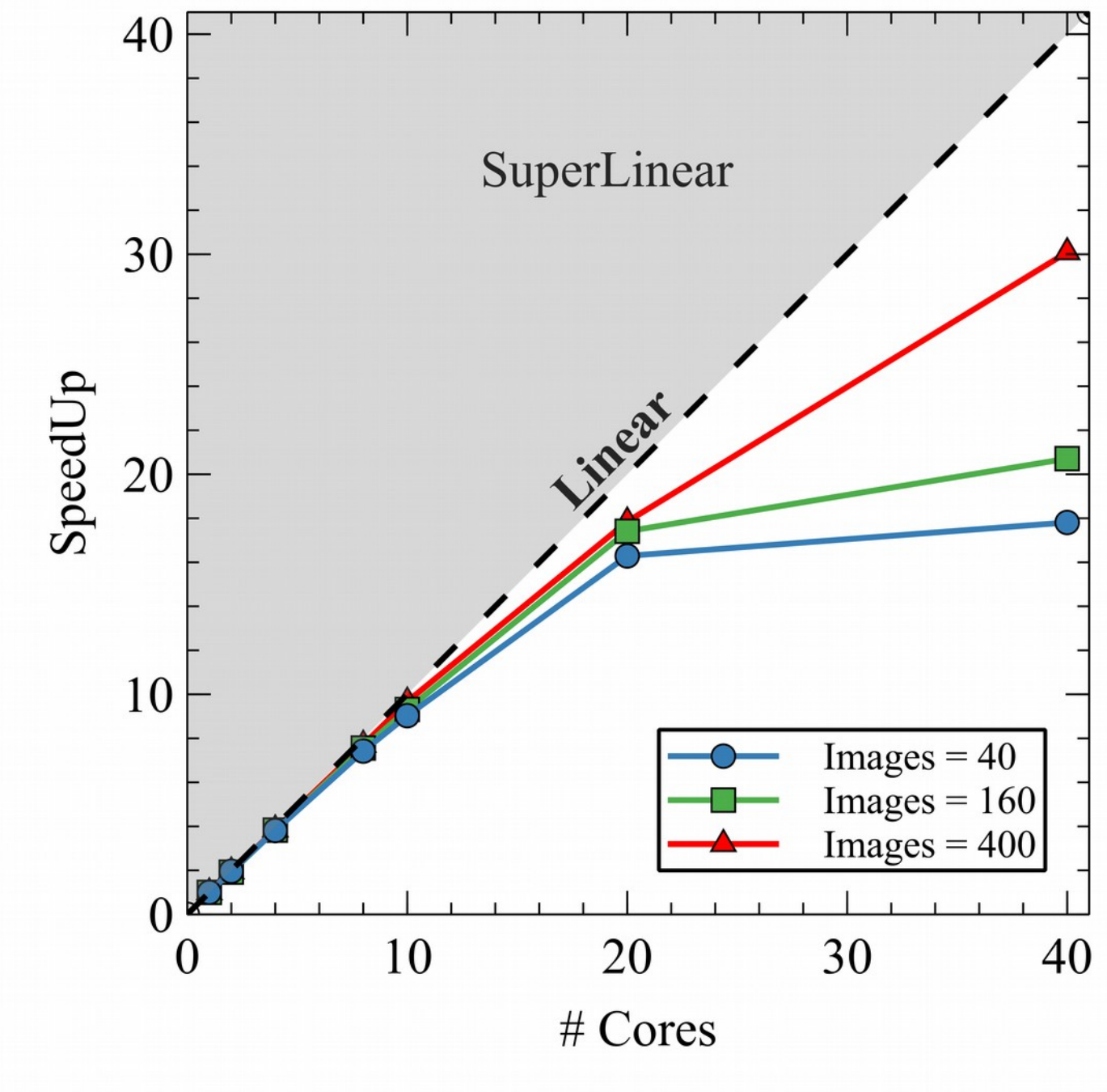

Figure 10: Speedup for different number of Images

4.2. Case Study

In order to demonstrate the method, we performed simulations for the small problem (one quarter of the pillar) for a pillar with the geometry of the inverted conical (frustum). The pillar is parameterized with respect to 4 variables *r, R, H* and *ρ* (Figure 11). The solution of the Eikonal equation takes only a few minutes, a number insignificant compared to the total time needed for the String Method to converge. The 50x50 mesh was used and the MEP was approximated with 40 Images. The simulations used 20 cores among 2 nodes, 10 cores on each node.

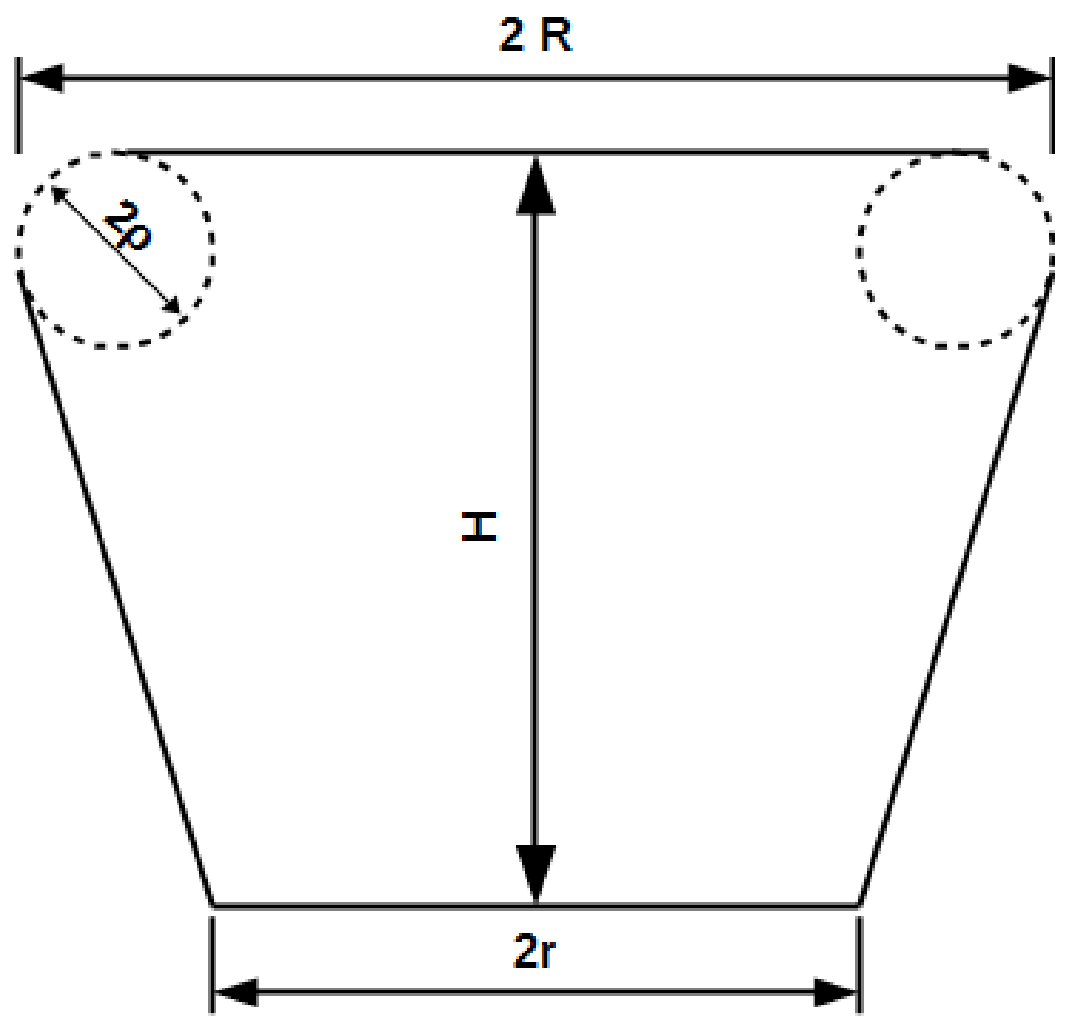

*Figure 11: Cross section of the inverted conical (frustum) with all the parameters visible*

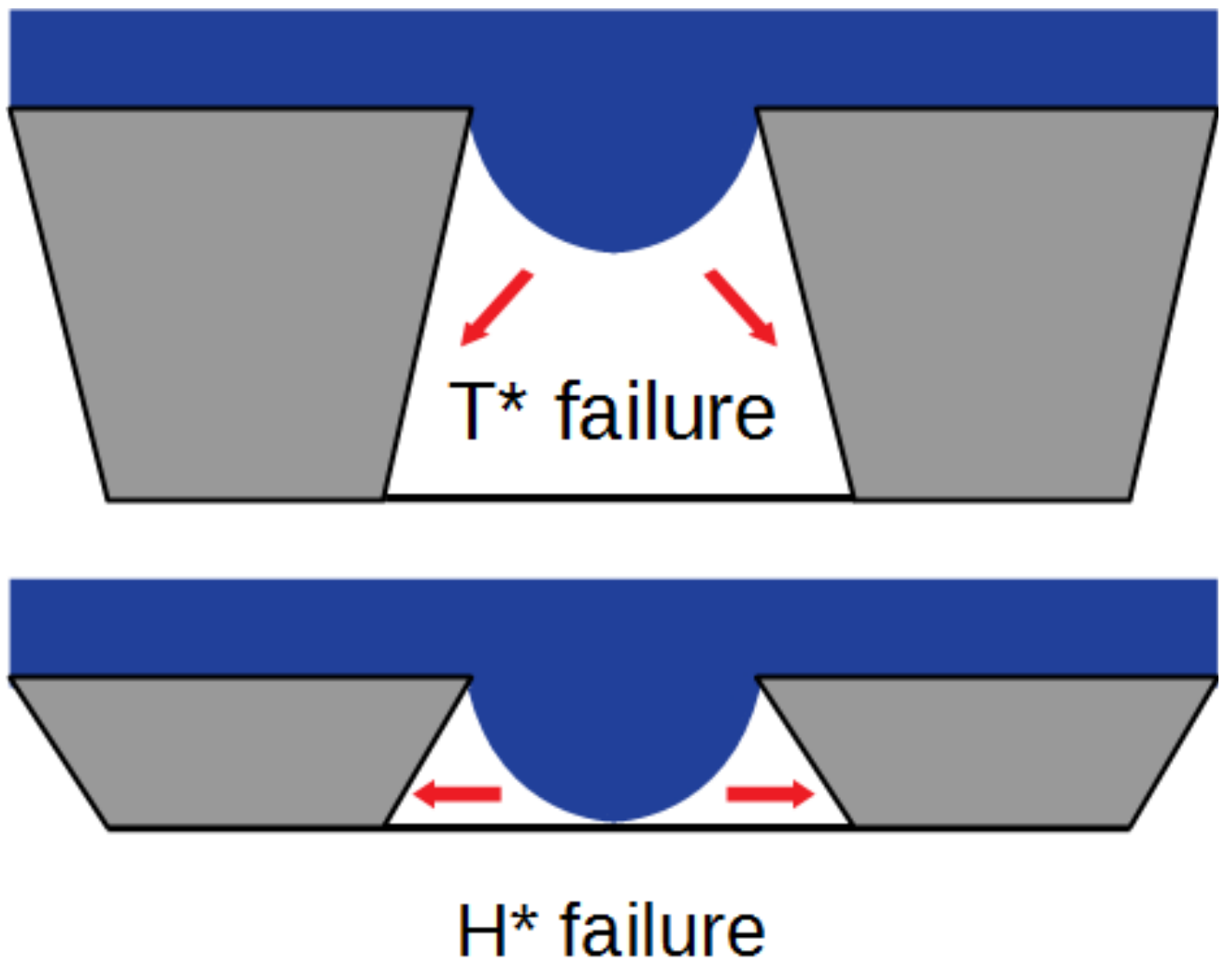

*Figure 12: Failure modes*

There are 2 major failure modes investigated in this paper (Figure 12). First one is called T* failure[39], according to which the liquid detaches from the edges of the pillar and wets the sidewalls and eventually collapses to the bottom. The greater the length of the sidewalls the more difficult for the interface to reach the bottom thus increasing the energy barrier. The second mode is called H* failure[39]. In this case the interface collapses to the bottom without wetting the sidewalls first. In this mode a lower energy barrier is expected.

First we examine the effect of the pillar base size on the barrier (Figure 13). The diagram reveals the existence of a maximum. The position of the maximum is the critical size where T* failure stops and H* failure begins. Specifically, small values of the radius should imply larger barrier because the size of the sidewall and thus the distance traveled by the liquid increases. That would be the case only if the liquid was able to follow the slope of the sidewall and collapse tangential to it (T* Failure), otherwise the interface would immediately collapse to the bottom of the geometry (H* Failure). The above mechanism described, implies that the maximum barrier corresponds to the smallest radius where T* Failure occurs, as confirmed by the calculations. Next, the effect of the radius of curvature of the top edge of the pillar, is examined (Figure 14). The higher the value of this radius, the smaller the barrier. High values imply a more blunt corner thus minimizing the resistance that the liquid faces towards its' descending, decreasing the barrier.

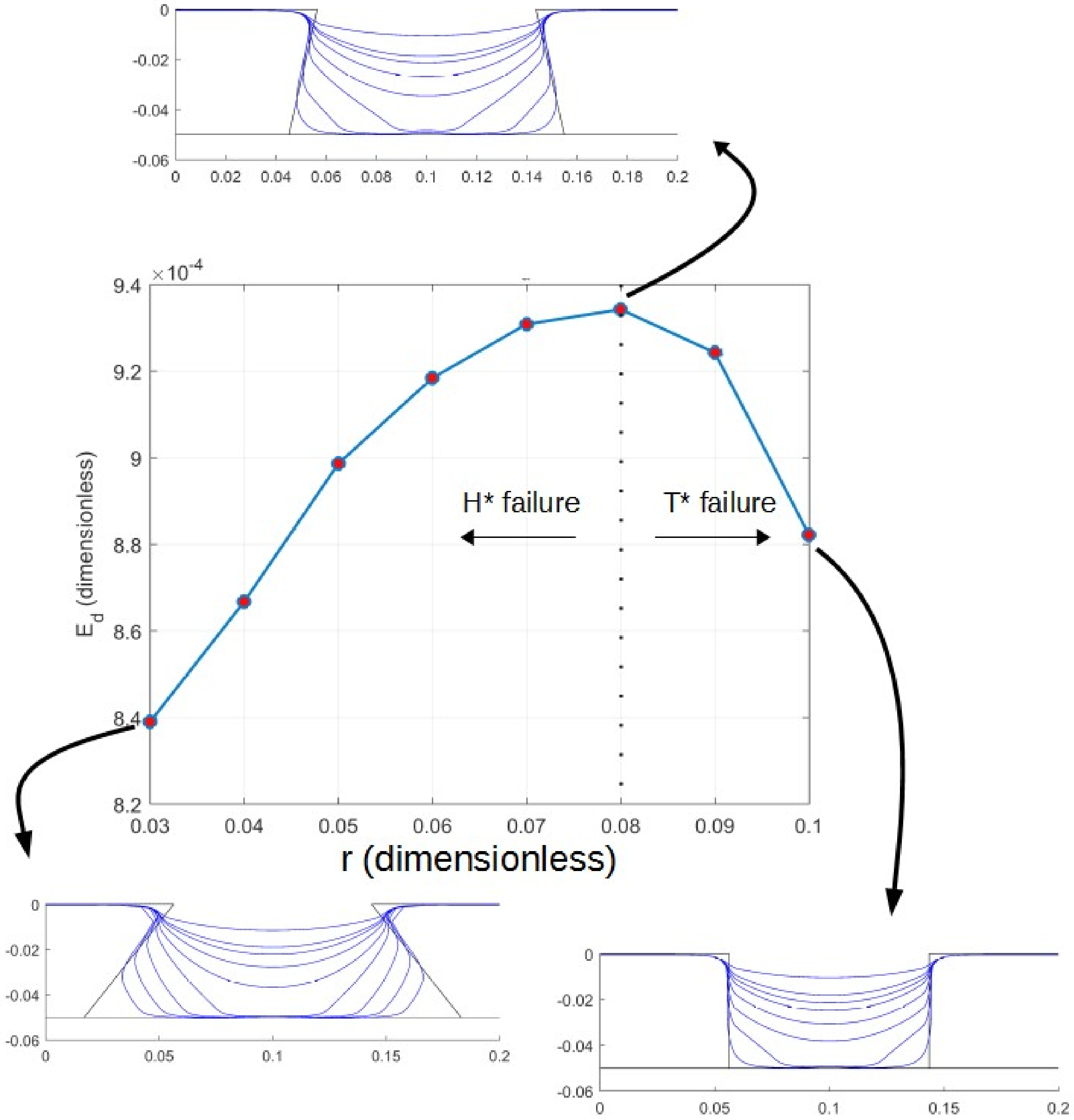

*Figure 13: Energy barrier against the base size "r". The arrows point at cross sections of the geometry. The blue lines correspond to the air-liquid interface while the system transitions from the CB to the W state.*

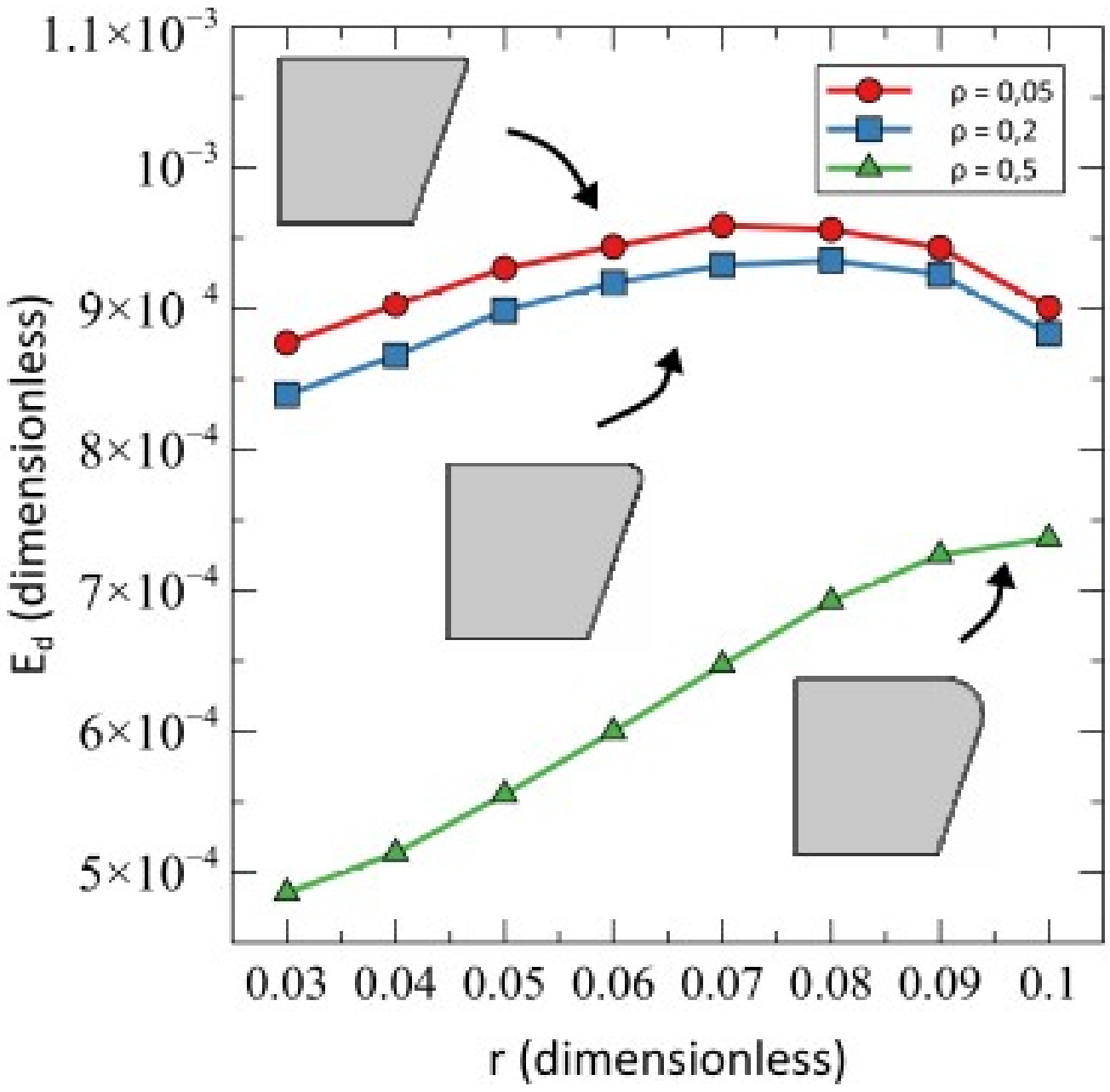

*Figure 14: Energy barrier of pillars with different radius of curvature against the base size "r".*

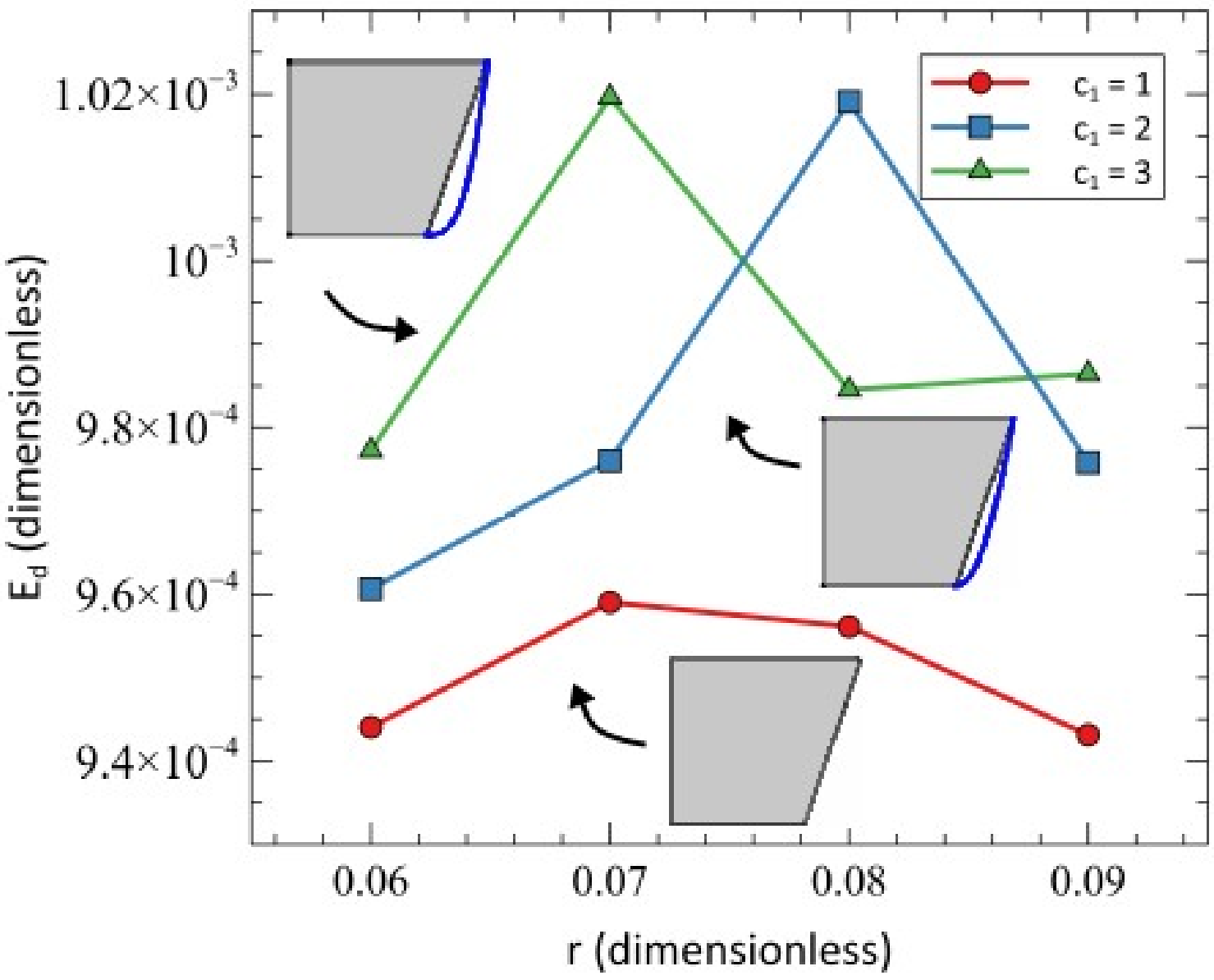

*Figure 15: Energy barrier of pillars with different "$c_1$" values against base size "r". The colored lines correspond to constant "$c_1$" values.*

Based on the above observations two modifications are proposed that are thought to produce better results (increased barrier), and will allow the investigation of more complex geometries. The two new geometries that are investigated are designed with parametric curves.

The first modification involves replacement of the side walls with a smooth curve (Figure 16), aiming to increase the length of the wall while keeping a slope small enough for T* Failure to take place. A right combination of the two is supposed to increase the distance traveled by the liquid and consequently the energy barrier of the transition. The curve is considered to be a parametric curve as described below:

$$s=[0 \quad 1] \tag{8.1}$$

$$x_1(s)=s(R-r) \tag{8.2}$$

$$y_1(s)=h\cdot s^{c_1} \tag{8.3}$$

Where "$c_1$" is a parameter adjusting the curvature. High values imply a longer curve, yet not a smooth slope.

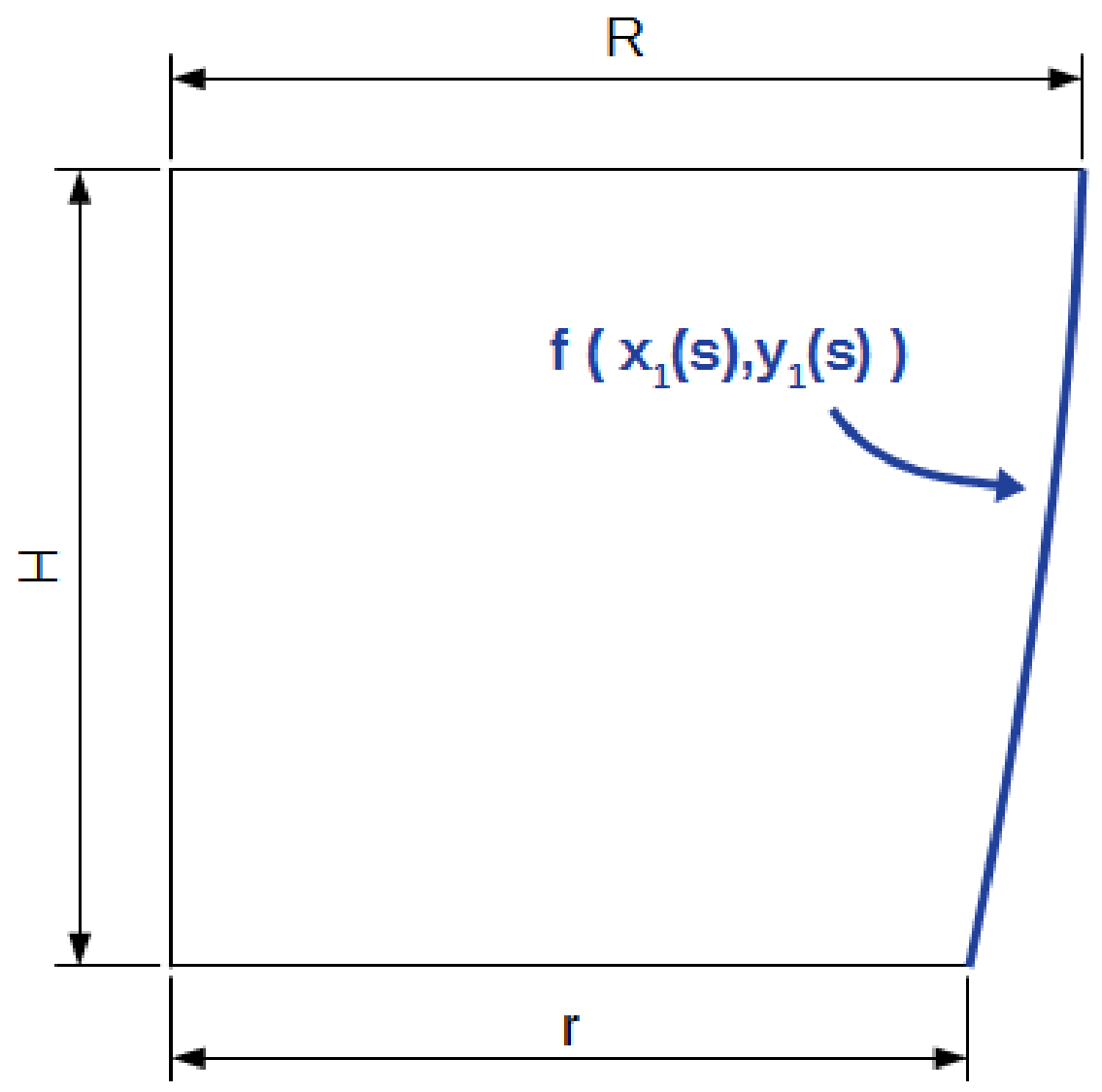

*Figure 17: First modification*

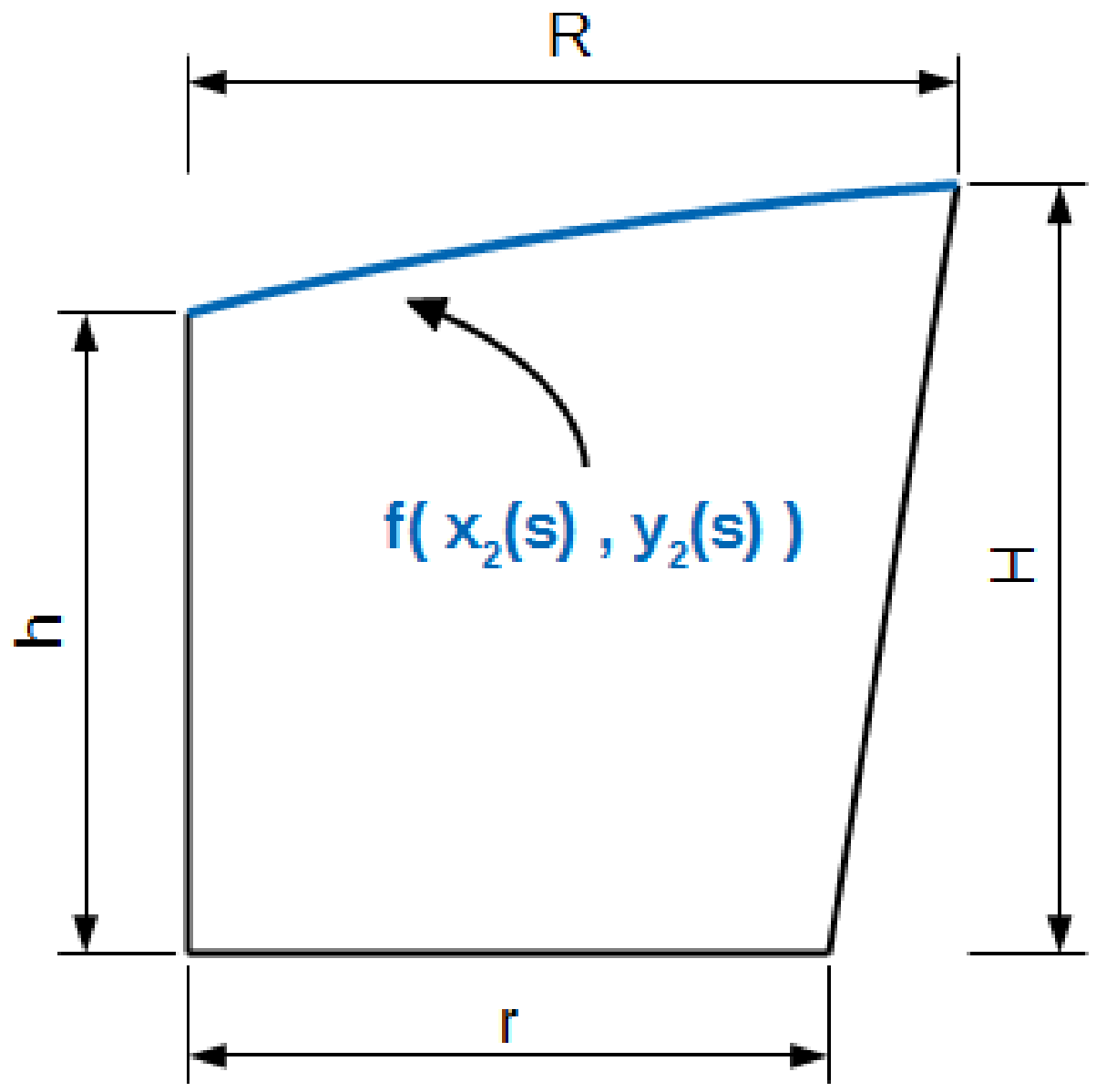

*Figure 16: Second modification*

This parametric study reveals a behavior similar to the original inverted conical (frustum), but with an increase of the energy barrier (Figure 15). The diagram simultaneously examines the effect of parameters "r" and "$c_1$". As can be observed, different combinations of the two may result in the same barrier. Based on the increased energy barrier the first modification is deemed a worth considering one, for the improvement of superhydrophobic behavior.

The second modification involves the replacement of the top wall, with a smooth curve, aiming at maximizing the top surface restraining the interface from collapsing (Figure 17). Initial assumptions indicate that the highest barrier correspond to pillars with the largest top surface area. The curve is considered to be a parametric curve as described below:

$$s=[0 \;\; 1] \tag{9.1}$$

$$x_2(s)=h s^{c_2} \tag{9.2}$$

$$y_2(s)=s(H-h) \tag{9.3}$$

Where "$c_2$" is a parameter adjusting the curvature. High values result in a pillar that approximates the original geometry.

The first parametric study conducted, assumed a fixed value for "h". Energy barrier is increased in cases where "$c_2$" value is not too small. The pillar corresponding to "$c_2$=1" is characterized by a sharp angle on the upper right corner. This angle favors the collapse of the interface, which having zero support from both sides easily collapses thus reducing the energy barrier of the transition.

The next parametric study takes under consideration the effect of parameter "h". As "h" decreases, the length of the side wall increases but the top right corner is getting sharper. As a result a critical "h" value exist, corresponding to a pillar with large upper surface and a smooth angle, yielding maximum energy barrier. The position of the maximum is independent of the small radius "r", as confirmed by the calculations. Barrier values generally surpass those of the original pillar.

Lastly the two modifications were combined, in order to study the total effect on the energy barrier. The simulations imply that the combination of the two modifications result in a barrier higher than each modification separately (Figure 18). Nevertheless, not every possible combination was tested.

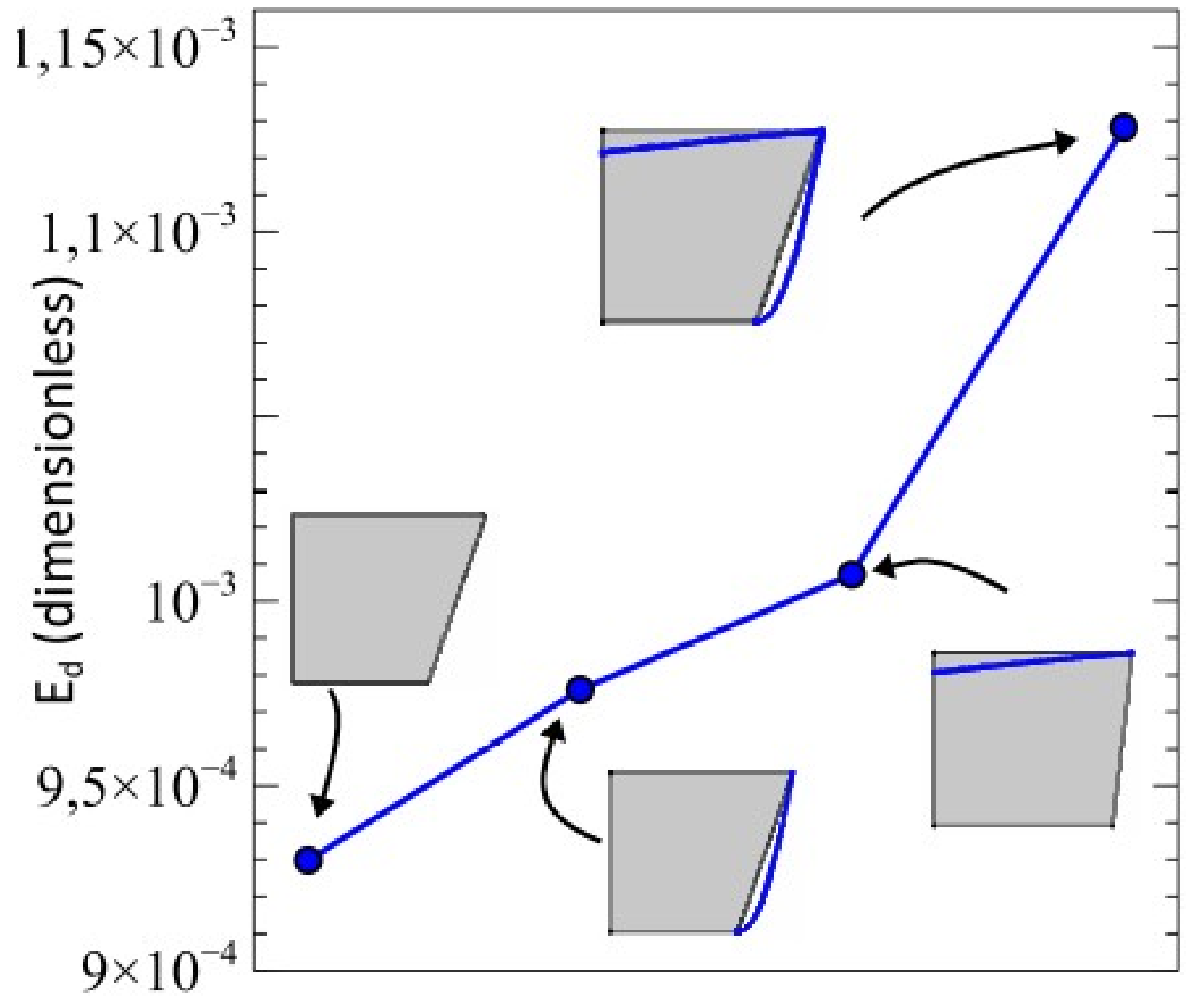

*Figure 18: Energy barrier of different geometries.*

5. Conclusions

The string method was tested for the case of the inverted conical (frustum). Firstly the original geometry was parameterized and a parametric investigation was performed. For the purpose of examining the finer geometric features of the pillar the small computational domain was used. Based on the results, the optimum geometric characteristics of the pillars were determined.

Moreover, we demonstrated the use of MEPs as a way to study the course of the air-solid interface collapse. Two mechanisms were analyzed and based on them, two modifications for the original geometry were proposed. These were tested and their positive contribution to the enhancement of the superhydrophobic behavior of the surface was confirmed. Lastly the two combinations were implemented simultaneously. Although not every possible set of parameters was tested, the results indicate enhanced superhydrophobic behavior.

In order to show that our method is capable not only to evaluate the suitability of pillars for the design of a superhydrophobic surface but also to provide results on a short time frame we conducted an analysis, based on the parallel speedup, to determine the scalability of the method. The results imply that our method is highly scalable and cost-effective, especially for the more computational intense cases (denser mesh and more Images for the representation of the MEP), because linear speedup can be observed even when using all available cores.